# Optical second harmonic generation in Yttrium Aluminum Borate single crystals (theoretical simulation and experiment)

Ali H Reshak*[1], S Auluck[2], A Majchrowski[3] and IV Kityk[4]

Address: [1]Institute of Physical Biology-South Bohemia University, Institute of System Biology and Ecology-Academy of Sciences – Nove Hrady 37333, Czech Republic, [2]Physics Department, Indian Institute of Technology, Kanpur (UP) 247667, India, [3]Institute of Applied Physics, Military University of Technology, Kaliskiego 2, 00-908 Warsaw, Poland and [4]Institue of Physics, J. Dlugosz University Czestochowa, Al. Armii Krajowej 13/15, Czestochowa, Poland, and Department of Chemistry, Silesian University of Technology, ul. Marcina Strzody 9, PL-44100 Gliwice, Poland

Email: Ali H Reshak* - maalidph@yahoo.co.uk; S Auluck - sauluck@gmail.com; A Majchrowski - i.kityk@ajd.czest.pl; IV Kityk - i.kityk@ajd.czest.pl

* Corresponding author

## Abstract

Experimental measurements of the second order susceptibilities for the second harmonic generation are reported for $YAl_3(BO_3)_4$ (YAB) single crystals for the two principal tensor components xyz and yyy. First principle's calculation of the linear and nonlinear optical susceptibilities for Yttrium Aluminum Borate $YAl_3(BO_3)_4$ (YAB) crystal have been carried out within a framework of the full-potential linear augmented plane wave (FP-LAPW) method. Our calculations show a large anisotropy of the linear and nonlinear optical susceptibilities. The observed dependences of the second order susceptibilities for the static frequency limit and for the frequency may be a consequence of different contribution of electron-phonon interactions. The imaginary parts of the second order SHG susceptibility $\chi^{(2)}_{123}(\omega)$, $\chi^{(2)}_{112}(\omega)$, $\chi^{(2)}_{222}(\omega)$, and $\chi^{(2)}_{213}(\omega)$ are evaluated. We find that the $2\omega$ inter-band and intra-band contributions to the real and imaginary parts of $\chi^{(2)}_{ijk}(\omega)$ show opposite signs. The calculated second order susceptibilities are in reasonable good agreement with the experimental measurements.

## I. Introduction

Yttrium Aluminium Borate $YAl_3(BO_3)_4$ (YAB) belongs to a family of double borates which crystallize in the trigonal structure of the mineral huntite $CaMg_3(CO_3)_4$ and belong to the space group R32 [1]. The general formula of these compounds is $RX_3(BO_3)_4$, where R = $Y^{3+}$, $Gd^{3+}$ or any other lanthanide, and X = $Al^{3+}$, $Sc^{3+}$, $Ga^{3+}$, $Cr^{3+}$, $Fe^{3+}$ [2]. YAB is a non centro-symmetric crystal and as early as in 1974 it was reported to be a very effective second-harmonic generating material [3]. Furthermore, owing to its good chemical stability and possibility of substituting $Y^{3+}$ ions with other lanthanide ions, namely $Nd^{3+}$, $Yb^{3+}$, $Dy^{3+}$ and $Er^{3+}$ [4] it is a promising material for laser applications. The nonlinear optical properties of this material in connection with lasing properties led to the construction of numerous systems generating red, green and blue light by self-frequency doubling effect [5]. YAB can be obtained as nano crystallite powders by simple technological approaches [6,7]. They also possess relatively large two-photon absorption [6], which makes them promising third order optical materials. At the same time they are good matrices for different rare earth ions [8-11]. The existing data is restrained by consideration of the local crystalline fields and the influence of the rare earth ions [12-15]. Intrinsic defects also may play an important role in determining the optical susceptibilities [16,17].

We feel that a reliable band structure will be of immense help in understanding the linear and nonlinear optical properties and show directions for technologists to obtain crystalline materials with desired optical properties. A reliable band structure calculation can help in determining the role of inter-band dipole matrix elements on the optical properties. It can give information on the dispersion of the bands in **k**-space and origin of the bands which are directly connected with the optical hyperpolarizabilities and susceptibilities [18]. In the present work we report first principle's calculation of the linear and nonlinear optical susceptibilities for YAB using the state-of-the-art full potential linear augmented plane wave method [19] which has proven to be one of the most accurate methods [20,21] for the computation of the electronic structure of solids within density functional theory (DFT) [22]. One specific feature of the borate crystals is the coexistence of the strong covalent and ionic chemical bonds, which provide relatively flat **k**-dispersion of the bands [23]. Moreover there exists substantial anisotropy of the chemical bonds which substantially restrains the application of the pseudopotential method, even norm-conserving one [24]. The aim of this paper is to understand the origin of birefringence and the high $\chi^2(\omega)$, using first principle's calculations.

In the Section 2 we present the computational and experimental details. Section 3 gives the results of the calculations and the measurements. The observed discrepancies are discussed following the band energy approach.

## II. Computational and experimental details

YAB has the trigonal structure (Fig. 1) of the mineral huntite [1]. YAB lattice provides suitable sites for rare-earth or transition metal ions doping [8]. The lattice parameters of YAB crystal are a = b = 9.295 Å, c = 7.243 Å, $\alpha = \gamma = 90°$ and $\beta = 120°$. YAB melts incongruently at 1280°C and decomposes into $YBO_3$ and $AlBO_3$. Therefore, it cannot be crystallized from stoichiometric melts.

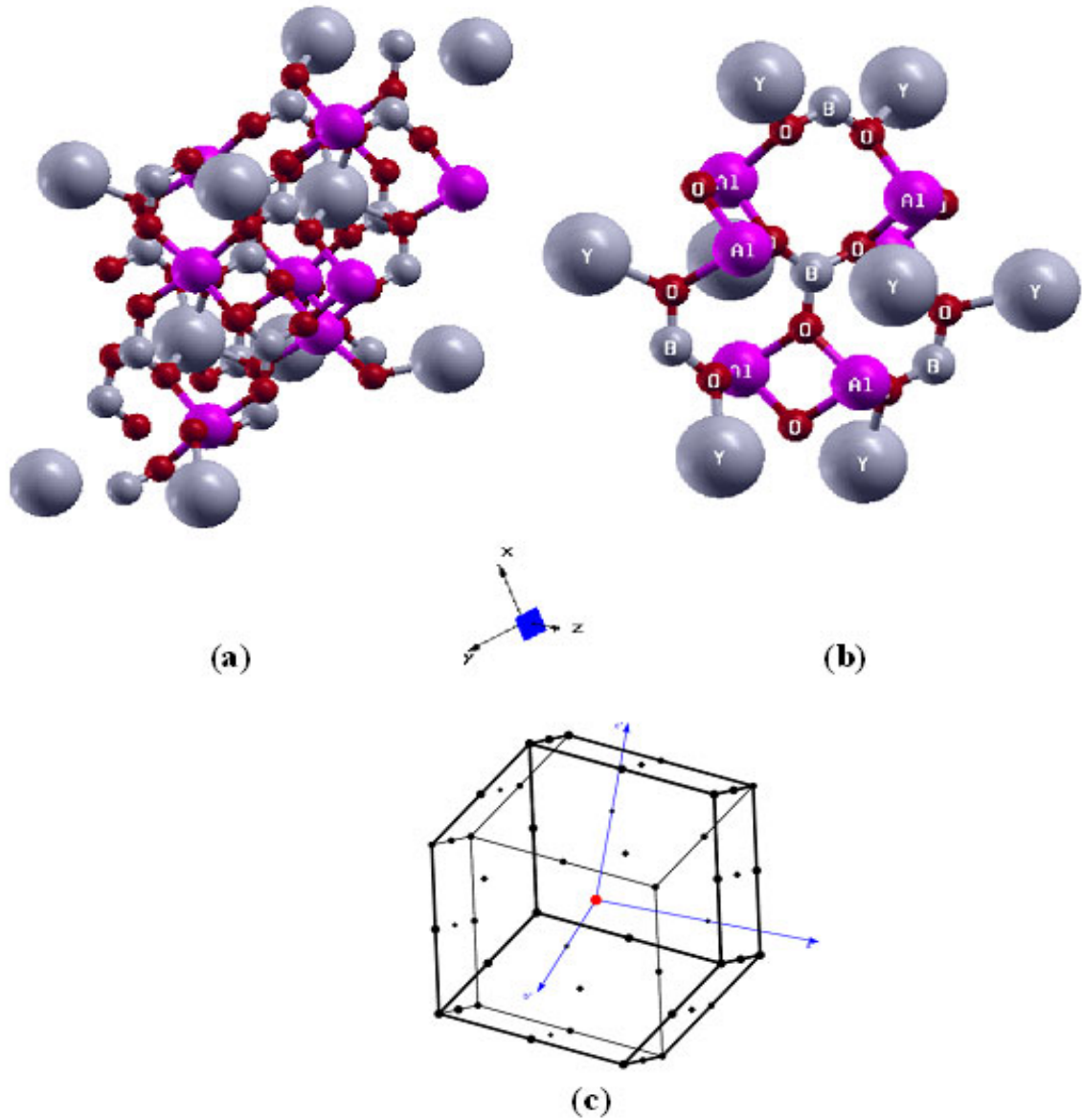

**Figure 1**
(a). Yttrium Aluminium Borate $YAl_3(BO_3)_4$ (YAB) crystal structure (b) Primitive unit cell, (c). Brillouin zone.

The application of high temperature solution growth (HTSG) method allows lowering the temperature of YAB crystallization below the temperature of the peritectic transformation.

Self-consistent calculations of the electronic structure and optical properties based on the scalar relativistic full-potential linearized augmented plane wave method were carried out using the WIEN2K package [19]. This is a very accurate and efficient scheme to solve the Kohn-Sham equation of density functional theory (DFT) [22]. This is an implementation of the DFT with different possible approximations for the exchange-correlation (XC) potential. The XC is treated within the local density approximation (LDA) [25] and scalar relativistic equations are used to obtain self-consistency. The Kohn-Sham equations are solved using a basis of linear APW's. In the interstitial region the potential and the charge density are represented by Fourier series. In order to achieve energy convergence, the wave functions in the interstitial region were expanded in plane waves with a cut-off $K_{max} = 9/R_{MT}$, where $R_{MT}$ denotes the smallest atomic sphere radius and $K_{max}$ gives the magnitude of the largest $K$ vector in the plane wave expansion. The $R_{MT}$ are taken to be 2.14, 1.81, 1.28, and 1.28 atomic units (a.u.) for Y, Al, B and O respectively. The valence wave functions inside the spheres are expanded up to $l_{max} = 10$ while the charge density was Fourier expanded up to $G_{max} = 14$.

Self-consistency is obtained using 200 $\vec{k}$ points in the irreducible Brillouin zone (IBZ). We calculated the frequency dependent linear optical properties using 500 $\vec{k}$ points and nonlinear optical properties using 1400 $\vec{k}$ points in the IBZ. The self-consistent calculations are assumed to be converged when the total energy of the system is stable within $10^{-5}$ Ry.

The second order optical susceptibilities were measured by standard method (Fig. 2) using 10 ns Nd-YAG laser (Carat, Lviv, Ukraine, 2005) at the fundamental wavelength 1064 nm, with pulse repetition 7 Hz. The Glan polarizers were used for definition of the input and output directions to measure the different tensor components of the second order optical susceptibilities. The green interferometric filter was used to cut the output doubled frequency signal at 533 nm with respect to the fundamental ones. Detection was performed by fast response photodiodes connected with the GHz oscilloscope (NewPort). The crystals were cut in directions which allowed to do the measurements for two principal tensor components $\chi^{(2)}_{123}(\omega)$ and $\chi^{(2)}_{222}(\omega)$. The set-up allows us to achieve a precision of 0.08 pm/V for the second order susceptibility.

## III. Results and Discussion
### A. First order optical susceptibilities and birefringence

We first consider the linear optical properties of the YAB crystal. The investigated crystals have trigonal symmetry, so we need to calculate two dielectric tensor components, corresponding to electric field $\vec{E}$ perpendicular and parallel to the optical c-axis, to completely characterize the linear optical properties. These are $\varepsilon^{xx}_2(\omega)$ and $\varepsilon^{zz}_2(\omega)$, the imaginary parts of the frequency dependent dielectric function. We have performed calculations of the frequency-dependent dielectric function using the expressions [31,32]

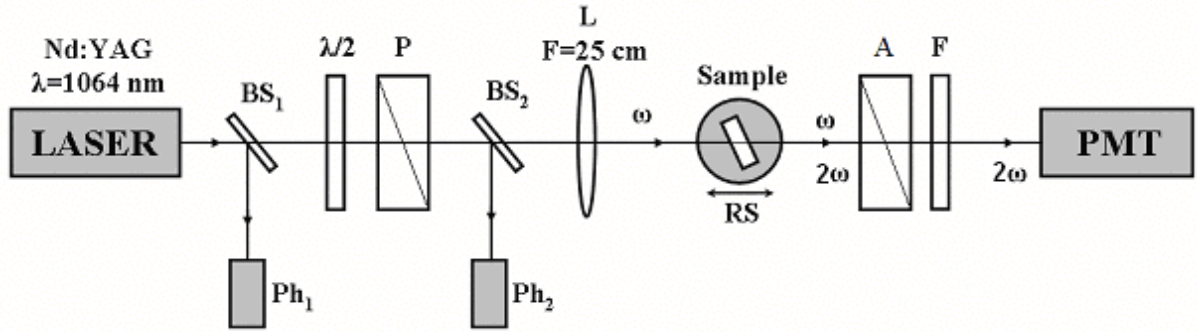

**Figure 2**
Second harmonic generation experimental setup (SHG): $BS_1$, $BS_2$-beam splitters, $Ph_1$, $Ph_2$-photodiodes, $\lambda/2$-half wave plate, P-Glan polarizer, A-Glan analyser, L-lens, RS-rotation stage, F-filter/s, PMT-photomultiplier tube.

$$\varepsilon_2^{zz}(\omega) = \frac{12}{m\omega^2} \int_{BZ} \sum \frac{\left|P_{nn'}^{Z}(k)\right|^2 dS_k}{\nabla \omega_{nn'}(k)} \quad (1)$$

$$\varepsilon_2^{xx}(\omega) = \frac{6}{m\omega^2} \int_{BZ} \sum \frac{\left[\left|P_{nn'}^{X}(k)\right|^2 + \left|P_{nn'}^{Y}(k)\right|^2\right] dS_k}{\nabla \omega_{nn'}(k)} \quad (2)$$

The above expressions are written in atomic units with $e^2 = 1/m = 2$ and $\hbar = 1$. where $\omega$ is the photon energy and $P_{nn'}^{X}(k)$ is the x component of the dipolar matrix elements between initial $|nk\rangle$ and final $|n'k\rangle$ states with their eigen-values $E_n(k)$ and $E_{n'}(k)$, respectively. $\omega_{nn'}(k)$ is the band energy difference $\omega_{nn'}(k) = E_n(k) - E_{n'}(k)$ and $S_k$ is a constant energy surface $S_k = \{k; \omega_{nn'}(k) = \omega\}$.

Figure 3 shows the calculated imaginary part of the anisotropic frequency dependent dielectric function $\varepsilon_2^{xx}(\omega)$ and $\varepsilon_2^{zz}(\omega)$. Broadening is taken to be 0.04 eV. All the optical properties are scissors corrected [26] using a scissors correction of 0.6 eV. This value is the difference between the calculated (5.1 eV) and measured energy gap (5.7 eV). The calculated energy gap is smaller than the experimental gap as expected from an LDA calculation [27]. It is well known that LDA calculations underestimate the energy gaps. A very simple way to overcome this drawback is to use the scissor correction, which merely makes the calculated energy gap equal to the experimental gap.

From Figure 3, it can be seen that the optical absorption edges for $\varepsilon_2^{xx}(\omega)$ and $\varepsilon_2^{zz}(\omega)$ are located at 5.7 eV. Thereafter a small hump arises at around 6.0 eV. Looking at these spectra we note that $\varepsilon_2^{xx}(\omega)$ and $\varepsilon_2^{zz}(\omega)$ increases to reach the highest magnitude at around 7.5 eV for $\varepsilon_2^{xx}(\omega)$, and around 8.5 eV for $\varepsilon_2^{zz}(\omega)$. It is known that peaks in the optical response are determined by the electric-dipole transitions between the valence and conduction bands. These peaks can be identified from the band structure. The calculated band structure along certain symmetry directions is given in Figure 4. In order to identify these peaks we need to look at the optical transition dipole matrix elements. We mark the transitions, giving the major structure in $\varepsilon_2^{xx}(\omega)$ and $\varepsilon_2^{zz}(\omega)$ in the band structure diagram. These transitions are labeled according to the spectral peak positions in Figure 3. For simplicity we have labeled the transitions in Figure 4, as A, B, and C. The transitions (A) are responsible for the structures of $\varepsilon_2^{xx}(\omega)$ and $\varepsilon_2^{zz}(\omega)$ in the energy range 0.0–5.0 eV, the transitions (B) 5.0–10.0 eV, and the transitions (C) 10.0–14.0 eV.

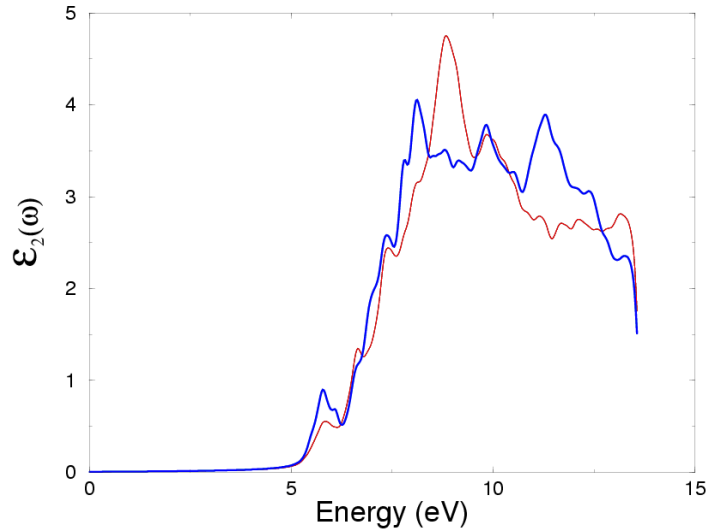

**Figure 3**
Calculated $\varepsilon_2^\perp(\omega)$ (dark curve) and $\varepsilon_2^\parallel(\omega)$ (light curve).

From the imaginary part of the dielectric function $\varepsilon_2^{xx}(\omega)$ and $\varepsilon_2^{zz}(\omega)$ the real part $\varepsilon_1^{xx}(\omega)$ and $\varepsilon_1^{zz}(\omega)$ is calculated by using of Kramers-Kronig relations [28]. The results of our calculated $\varepsilon_1^{xx}(\omega)$ and $\varepsilon_1^{zz}(\omega)$ are shown in Figure 5. The calculated $\varepsilon_1^{xx}(\omega)$ is about 2.4 and $\varepsilon_1^{zz}(\omega)$ is about 2.5.

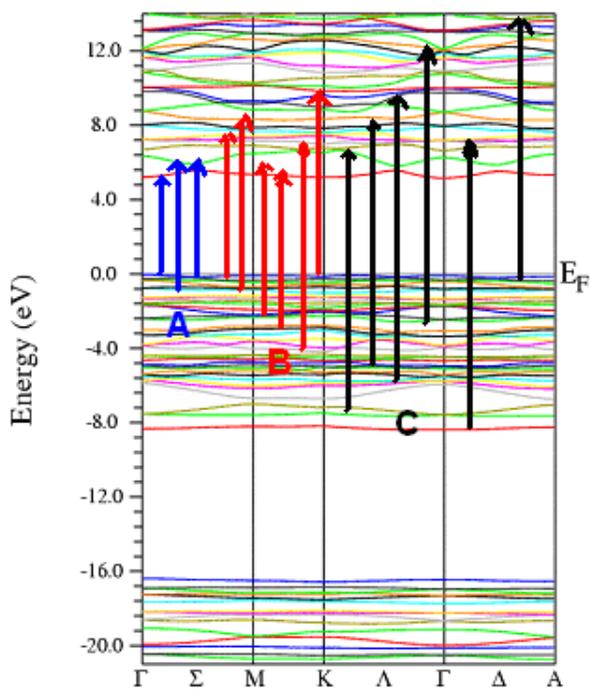

**Figure 4**
The optical transitions shown on the band structure of YAB.

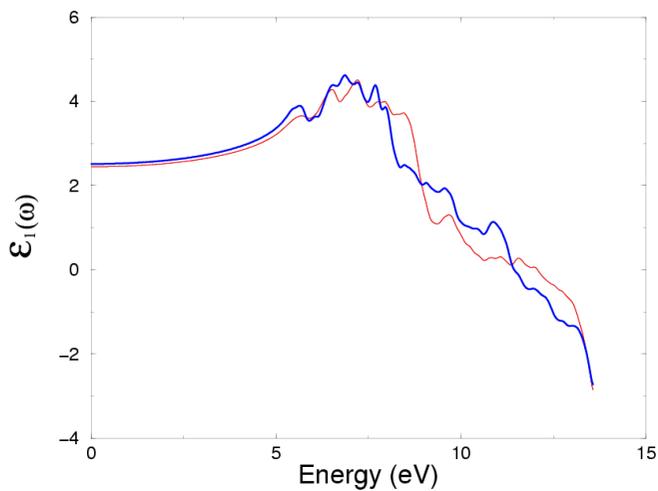

**Figure 5**
Calculated $\varepsilon_1^\perp(\omega)$ (dark curve) and $\varepsilon_1^{||}(\omega)$ (light curve).

These crystals show considerable anisotropy in the linear optical susceptibilities which favors large SHG susceptibilities. The birefringence is also important in fulfilling phase-matching conditions. The birefringence can be calculated from the linear response functions from which the anisotropy of the index of refraction is obtained. One can determine the value of the extraordinary and ordinary refraction indices. The birefringence is a difference between the extraordinary and ordinary refraction indices, $\Delta n = n_e - n_0$, where $n_e$ is the index of refraction for an electric field oriented along the **c**-axis and $n_0$ is the index of refraction for an electric field perpendicular to the **c**-axis. Figure 6, shows the birefringence $\Delta n(\omega)$ for this single crystal. The birefringence is important only in the non-absorbing region, which is below the energy gap. In the absorption region, the absorption will make it difficult for these compounds to be used as nonlinear crystals in optic parametric oscillators or frequency doublers and triplers. We note that the spectral feature of $\Delta n(\omega)$ shows strong oscillations around zero in the energy range up to 12.5 eV. Thereafter it drops to zero. We find that the calculated birefringence at zero energy is 0.025 in excellent agreement with our own measurement of 0.02. It is known that for the borates, the contribution of the electron-phonon interaction to the dielectric dispersion may be neglected for the SHG effects [29] contrary to the linear electro-optics Pockels effect. Comparing these dependences with the anisotropy for other borates [23,24] one can conclude that the anisotropy caused by the chemical bonds is smaller than in the other borates [24].

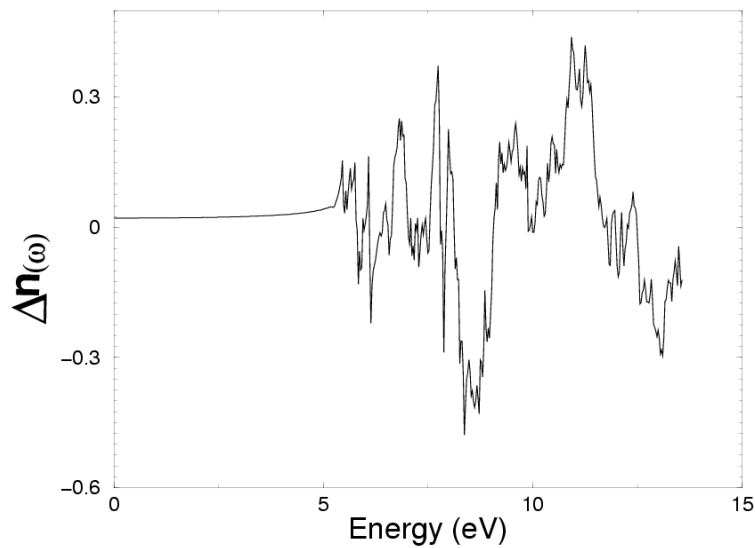

**Figure 6**
Calculated $\Delta n(\omega)$.

### B. Second order susceptibilities

The expressions of the complex second-order nonlinear optical susceptibility tensor $\chi_{ijk}^{(2)}(-2\omega;\omega;\omega)$ has been presented in previous works [33,34]. From the expressions we can obtain the three major contributions: the interband transitions $\chi_{\text{inter}}^{ijk}(-2\omega;\omega,\omega)$, the intraband transitions $\chi_{\text{intra}}^{ijk}(-2\omega;\omega,\omega)$ and the modulation of interband terms by intraband terms $\chi_{\text{mod}}^{ijk}(-2\omega;\omega,\omega)$. These are

$$\chi_{\text{inter}}^{ijk}(-2\omega;\omega,\omega) = \frac{e^3}{\hbar^2} \sum_{nml} \int \frac{d\vec{k}}{4\pi^3} \frac{\vec{r}_{nm}^i \{\vec{r}_{ml}^j \vec{r}_{ln}^k\}}{(\omega_{ln}-\omega_{ml})} \left\{ \frac{2f_{nm}}{(\omega_{mn}-2\omega)} + \frac{f_{ml}}{(\omega_{ml}-\omega)} + \frac{f_{ln}}{(\omega_{ln}-\omega)} \right\} \tag{3}$$

$$\chi_{\text{intra}}^{ijk}(-2\omega;\omega,\omega) = \frac{e^3}{\hbar^2} \int \frac{d\vec{k}}{4\pi^3} \left[ \sum_{nml} \omega_{nm} \vec{r}_{nm}^i \{\vec{r}_{ml}^j \vec{r}_{ln}^k\} \left\{ \frac{f_{nl}}{\omega_{ln}^2(\omega_{ln}-\omega)} - \frac{f_{lm}}{\omega_{ml}^2(\omega_{ml}-\omega)} \right\} \right.$$
$$\left. -8i \sum_{nm} \frac{f_{nm} \vec{r}_{nm}^i \{\Delta_{mn}^j \vec{r}_{nm}^k\}}{\omega_{mn}^2(\omega_{mn}-2\omega)} + 2\sum_{nml} \frac{f_{nm} \vec{r}_{nm}^i \{\vec{r}_{ml}^j \vec{r}_{ln}^k\}(\omega_{ml}-\omega_{ln})}{\omega_{mn}^2(\omega_{mn}-2\omega)} \right] \tag{4}$$

$$\chi_{\text{mod}}^{ijk}(-2\omega;\omega,\omega) = \frac{e^3}{2\hbar^2} \int \frac{d\vec{k}}{4\pi^3} \left[ \sum_{nml} \frac{f_{nm}}{\omega_{mn}^2(\omega_{mn}-\omega)} \left\{ \omega_{nl} \vec{r}_{lm}^i \{\vec{r}_{mn}^j \vec{r}_{nl}^k\} - \omega_{lm} \vec{r}_{nl}^i \{\vec{r}_{lm}^j \vec{r}_{mn}^k\} \right\} \right.$$
$$\left. -i \sum_{nm} \frac{f_{nm} \vec{r}_{nm}^i \{\vec{r}_{mn}^j \Delta_{mn}^k\}}{\omega_{mn}^2(\omega_{mn}-\omega)} \right] \tag{5}$$

where $n \neq m \neq l$. Here $n$ denotes the valence states, $m$ the conduction states and $l$ denotes all states ($l \neq m, n$). There are two kinds of transitions which take place one of them $vcc'$, involving one valence band ($v$) and two conduction bands ($c$ and $c'$), and the second transition $vv'c$, involving two valence bands ($v$ and $v'$) and one conduction band ($c$). The symbols are defined as $\Delta_{nm}^i(\vec{k}) = \vartheta_{nn}^i(\vec{k}) - \vartheta_{mm}^i(\vec{k})$ with $\vec{\vartheta}_{nm}^i$ being the $i$ component of the electron velocity given as $\vartheta_{nm}^i(\vec{k}) = i\omega_{nm}(\vec{k}) r_{nm}^i(\vec{k})$ and $\{r_{nm}^i(\vec{k}) r_{ml}^j(\vec{k})\} = \frac{1}{2}(r_{nm}^i(\vec{k}) r_{ml}^j(\vec{k}) + r_{nm}^j(\vec{k}) r_{ml}^i(\vec{k}))$. The position matrix elements between states *n* and *m*, $r^i_{nm}(\vec{k})$, are calculated from the momentum matrix element $P^i_{nm}$ using the relation [35]: $r^i_{nm}(\vec{k}) = \frac{P^i_{nm}(\vec{k})}{im\omega_{nm}(\vec{k})}$, with the energy difference between the states *n* and *m* given by $\hbar\omega_{nm} = \hbar(\omega_n - \omega_m)$. $f_{nm} = f_n - f_m$ is the difference of the Fermi distribution functions. *i*, *j* and *k* correspond to cartesian indices.

It has been demonstrated by Aspnes [36] that only one virtual-electron transitions (transitions between one valence band state and two conduction band states) give a significant contribution to the second-order tensor. Hence we ignore the virtual-hole contribution (transitions between two valence band states and one conduction band state) because it was found to be negative and more than an order of magnitude smaller than the virtual-electron contribution for these compounds. For simplicity we denote $\chi^{(2)}_{ijk}(-2\omega;\omega;\omega)$ by $\chi^{(2)}_{ijk}(\omega)$.

We have measured the second order susceptibilities of YAB single crystal using Nd-YAG laser at the fundamental wavelength 1064 nm. Since the investigated crystals belong to the point group R32 there are only five independent components of the SHG tensor, namely, the 123, 112, 222, 213 and 312 components (1, 2, and 3 refer to the x, y and z axes, respectively) [30]. These are $\chi^{(2)}_{123}(\omega)$, $\chi^{(2)}_{112}(\omega)$, $\chi^{(2)}_{222}(\omega)$, $\chi^{(2)}_{213}(\omega)$ and $\chi^{(2)}_{312}(\omega)$. Here $\chi^{(2)}_{ijk}(\omega)$ is the complex second-order nonlinear optical susceptibility tensor $\chi^{(2)}_{ijk}(-2\omega;\omega;\omega)$. The subscripts *i*, *j*, and *k* are Cartesian indices.

The calculated imaginary part of the second order SHG susceptibilities $\chi^{(2)}_{123}(\omega)$, $\chi^{(2)}_{112}(\omega)$, $\chi^{(2)}_{222}(\omega)$, and $\chi^{(2)}_{213}(\omega)$ are shown in Figures 7 and 8. We do not show the component $\chi^{(2)}_{312}(\omega)$ because it is very small. Our calculation and measurement show that $\chi^{(2)}_{222}(\omega)$ is the dominant component which shows the largest *total Re* $\chi^{(2)}_{ijk}(0)$ value compared to the other components (Table 1). A definite enhancement in the anisotropy on going from linear optical properties to the nonlinear optical properties is evident (Figures 7 and 8). It is well known that nonlinear optical susceptibilities are more sensitive to small changes in the band structure than the linear optical ones. Hence any anisotropy in the linear optical properties is enhanced more significantly in the nonlinear spectra.

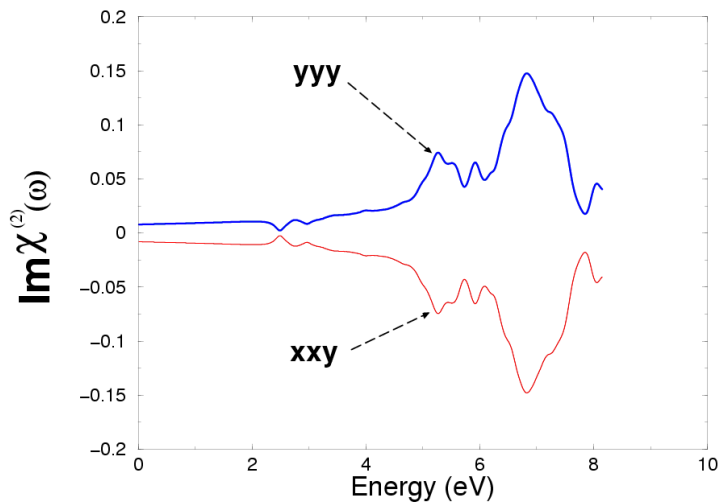

### Figure 7
Calculated Im $\chi^{(2)}_{222}(\omega)$ (dark curve) and Im $\chi^{(2)}_{112}(\omega)$ (light curve), all Im $\chi^{(2)}(\omega)$ are multiplied by $10^{-7}$, and in esu units.

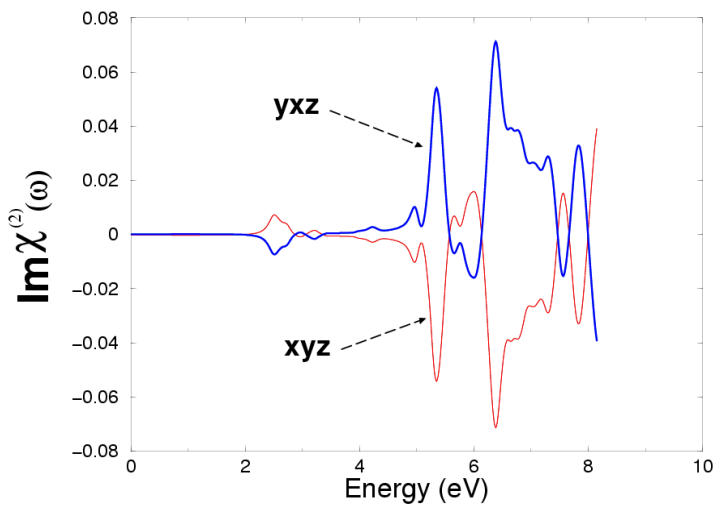

### Figure 8
Calculated Im $\chi^{(2)}_{213}(\omega)$ (dark curve) and Im $\chi^{(2)}_{123}(\omega)$ (light curve), all Im $\chi^{(2)}(\omega)$ are multiplied by $10^{-7}$, and in esu units.

**Table 1:** Calculated total, intra-band and inter-band contributions of *Re* $\chi^{(2)}_{ijk}(0)$ in units of $10^{-7}$ esu, along with the measured $\chi^{(2)}_{ijk}(0)$ in units of pm/V.

| Component | 123 | 112 | 222 | 213 |
| --- | --- | --- | --- | --- |
| Re $\chi^{ijk}$(0)total | -0.0011 | -0.009 | 0.01 | 0.0002 |
| Re $\chi^{ijk}$(0)int er | -0.007 | 0.035 | -0.038 | 0.006 |
| Re $\chi^{ijk}$(0)int ra | 0.008 | -0.04 | 0.04 | -0.0085 |
| Total Re $\chi^{ijk}$(0) pm/V | -0.5 | -0.7 | 0.8 | 0.5 |
| Experimental SHG $\chi^{ijk}$(0) (pm/V) | 0.12 | | 1.03 | |

In Figure 9, we show the $2\omega$ inter-band and intra-band contributions to Im $\chi^{(2)}_{222}(\omega)$. We note the opposite signs of the two contributions throughout the frequency range. We have calculated the total complex susceptibility for $\chi^{(2)}_{123}(\omega)$,... and $\chi^{(2)}_{213}(\omega)$. The real part of the dominant component is shown in Figure 10. The zero-frequency limit of all components is listed in Table 1.

From above we can see the total second order susceptibility determining SHG is zero below half the band gap. The $2\omega$ terms start contributing at energies $\sim 1/2E_g$ and the $\omega$ terms for energy values above $E_g$. In the low energy regime ($\leq 5$ eV) the SHG optical spectra is dominated by the

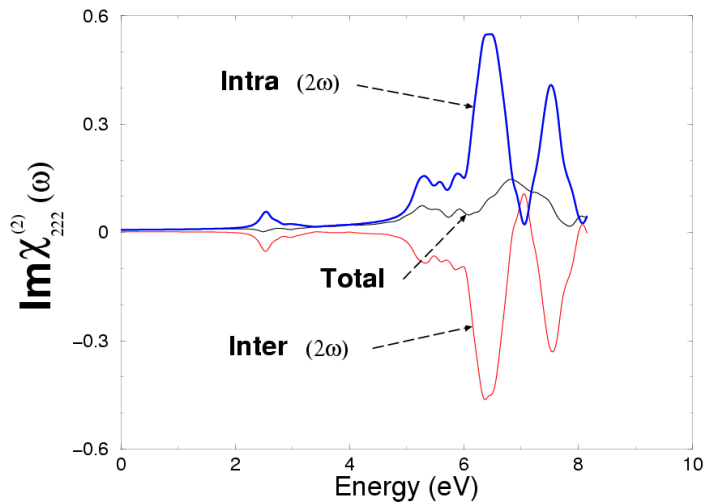

**Figure 9**
Calculated Im $\chi^{(2)}_{222}(\omega)$ along with the intra ($2\omega$) and inter ($2\omega$)-band contributions. All Im $\chi^{(2)}(\omega)$ are multiplied by $10^{-7}$, and in esu units.

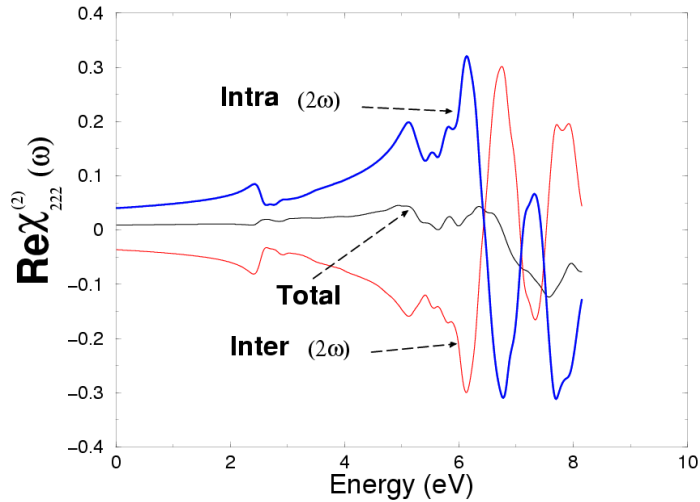

**Figure 10**

Calculated Re $\chi^{(2)}_{222}(\omega)$ along with the intra (2$\omega$) and inter (2$\omega$)-band contributions. All Im $\chi^{(2)}(\omega)$ are multiplied by $10^{-7}$, and in esu units.

2$\omega$ contributions. Beyond 5.7 eV (values of the fundamental energy gaps) the major contribution comes from the $\omega$ term.

One could expect that the structures in Im $\chi^{(2)}_{ijk}(\omega)$ could be understood from the features of $\varepsilon_2(\omega)$. Unlike the linear optical spectra, the features in the SHG susceptibility are very difficult to identify from the band structure because of the presence of 2$\omega$ and $\omega$ terms. But we use of the linear optical spectra to identify the different resonance leading to various features in the SHG spectra. The first spectral band in Im $\chi^{(2)}_{123}(\omega)$ between 0.0–5.0 eV is mainly originated from 2$\omega$ resonance and arises from the first structure in $\varepsilon_2(\omega)$. The second band between 5.0–7.0 eV is associated with interference between the $\omega$ resonance and 2$\omega$ resonance and is associated with high structure in $\varepsilon_2(\omega)$. The last structure from 7.0–8.0 eV is mainly due to $\omega$ resonance and is associated with the tail in $\varepsilon_2(\omega)$.

From an experimental viewpoint, one of the quantities of interest is the magnitude of SHG (proportional to the second order susceptibility). We present the absolute values of $\chi^{(2)}_{123}(\omega) = \chi^{(2)}_{213}(\omega)$, and $\chi^{(2)}_{112}(\omega) = \chi^{(2)}_{222}(\omega)$ in Figure 11. The first peak for these components are located at 2$\omega$ = 5.31 and 5.11 eV with the peak values of (0.052 and 0.081) × $10^{-7}$ esu, respectively. To evaluate the performed calculations we have done the measurements of the absolute value of $\chi^{(2)}_{123}(\omega)$ and $\chi^{(2)}_{222}(\omega)$ for the YAB single crystals for the Nd-YAG laser wavelength 1064 nm and we have revealed the corresponding values equal to about (0.042 and 0.061)×$10^{-7}$ esu, respectively confirming sufficiently good agreement. The calculated second order susceptibilities show substantially good agreement with the measured one.

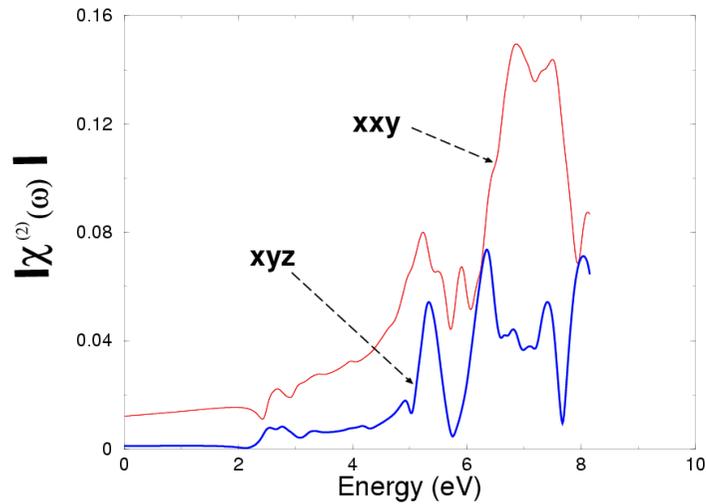

**Figure 11**
Calculated absolute value of $\chi^{(2)}_{123}(\omega)$ (dark curve) and $\chi^{(2)}_{112}(\omega)$ (light curve). All absolute values of $\chi^{(2)}(\omega)$ are multiplied by $10^{-7}$, and in esu units.

## IV. Conclusion

We have performed experimental measurements of the second order susceptibilities for the second harmonic generation for the $YAl_3(BO_3)_4$ (YAB) single crystals for the two principal tensor components xyz and yyy. We have reported a first principle's calculation of the linear and nonlinear optical susceptibilities using the FP-LAPW method within a framework of DFT. Our calculations show that YAB possesses a direct energy band gap of about 5.1 eV located at Γ point of the Brillouin zone. This is smaller than the experimental value of 5.7 eV. The calculated imaginary and real parts of the second order SHG susceptibility $\chi^{(2)}_{123}(\omega) = \chi^{(2)}_{213}(\omega)$ and $\chi^{(2)}_{222}(\omega) = \chi^{(2)}_{112}(\omega)$ were found to be in reasonable agreement with the measurements. We note that any anisotropy in the linear optical susceptibilities will significantly enhance the nonlinear optical susceptibilities. Our calculations show that the $2\omega$ inter-band and intra-band con-tributions to the real and imaginary parts of $\chi^{(2)}_{ijk}(\omega)$ show opposite signs. This fact may be used in future for molecular engineering of the crystals in the desirable directions.

**Acknowledgements**
The authors would like to thank the Institute of Physical Biology and Institute of System Biology and Ecology-Computer Center for providing the computational facilities. This work was supported from the institutional research concept of the Institute of Physical Biology, UFB (No. MSM6007665808), and the Institute of System Biology and Ecology, ASCR (No. AVOZ60870520).